\documentclass[fleqn,usenatbib]{mnras}

\usepackage{newtxtext,newtxmath}

\usepackage[T1]{fontenc}
\usepackage{ae,aecompl}

\usepackage{graphicx}	
\usepackage{amsmath}	
\usepackage{amssymb}	

\usepackage{etoolbox}
\makeatletter
\patchcmd\@combinedblfloats{\box\@outputbox}{\unvbox\@outputbox}{}{\errmessage{\noexpand patch failed}}
\makeatother

\title[Post-processing and Disruption of Substructure]{Leavers and remainers: Galaxies split by group-exit} 

\author[N. Choque-Challapa et al.]{Nelvy Choque-Challapa,$^{1}$\thanks{E-mail: n.c.choque@astro.rug.nl}
Rory Smith,$^{2}$
Graeme Candlish,$^{3}$
Reynier Peletier,$^{1}$
\newauthor
and Jihye Shin$^{2}$
\\
$^{1}$Kapteyn Astronomical Institute, University of Groningen, Landleven 12, NL-9747 AD Groningen, the Netherlands\\
$^{2}$Korea Astronomy \& Space Science Institute, Daejeon 305-348, Republic of Korea\\
$^{3}$Instituto de F\'isica y Astronom\'ia, Universidad de Valpara\'iso, Avda. Gran Breta\~na 1111 Valpara\'iso, Chile
}

\date{Accepted XXX. Received YYY; in original form ZZZ}

\pubyear{2019}

\begin{document}
\label{firstpage}
\pagerange{\pageref{firstpage}--\pageref{lastpage}}
\maketitle

\begin{abstract}
The disruption of substructure in galaxy clusters likely plays an important role in shaping the cluster population as a significant fraction of cluster galaxies today have spent time in a previous host system, and thus may have been pre-processed. Once inside the cluster, group galaxies face the combined environmental effects from group and cluster - so called `post-processing'. We investigate these concepts, by tracking the evolution of satellites and their hosts after entering the cluster and find that tidal forces during their first pericentric passage are very efficient at breaking up groups, preferentially removing satellites at larger distances from their hosts. 92.2\% of satellites whose host has passed pericentre will leave their host by $z=0$, typically no later than half a Gyr after pericentric passage. We find satellites leave with high velocities, and quickly separate to large distances from their hosts, making their identification within the cluster population challenging. Those few satellites ($\sim7.8\%$) that remain bound to their hosts after a pericentric passage are typically found close to their host centres. This implies that substructure seen in clusters today is very likely on first infall into the cluster, and yet to pass pericentre. This is even more likely if the substructure is extended, with satellites beyond R$_{200}$ of their host. We find the cluster dominates the tidal mass loss and destruction of satellites, and is responsible for rapidly halting the accretion of new satellites onto hosts once they reach 0.6-0.8 R$_{200}$ radii from the cluster.
\end{abstract}




\section{Introduction}
The importance of environmental effects for the evolution of galaxies in clusters has long been debated. Galaxies in cluster cores are more likely to be elliptical or spheroidal, in comparison with galaxies in the field \citep{Dressler80,Postman2005,Poggianti2008}, additionally showing older and redder stellar populations. In order to understand this environmental dependency, the combination of observations and simulations can give us more insights into galaxy evolution. This has led to the understanding that, in fact, galaxies in high-density environments may be morphologically transformed as they are exposed to interactions with other galaxies and the overall potential of the cluster \citep{Gnedin2003a,Gnedin2003b}, as well as the intra cluster/group medium \citep{Gunn1972}. Galaxies, and their dark matter halos, can experience strong tidal forces when they enter host groups and clusters, leading to tidal stripping of their mass \citep{Wetzel10}. Mass loss can also arise from high speed galaxy-galaxy encounters, a process known as ``harassment" \citep{Moore1996,Mastropietro2005,Smith2015}, or during galaxy-galaxy mergers \citep{Toomre1972,Angulo2009}. Additionally, the motion of a galaxy through the hot gas that fills the cluster potential well, known as the intra-cluster medium (or intra-group medium in galaxy groups), can remove the galaxy's cold, atomic, disk gas in a process known as ram pressure stripping \citep{Gunn1972,Roediger2007,Vijayaraghavan2015}.

The evidence, however, that some cluster galaxies may have been strongly influenced in the galaxy group environment, prior to falling into the cluster (referred to as \textit{pre-processing}, \citealp{Mihos2004}) has increased substantially. A number of studies have quantified the fraction of galaxies in clusters that have previously spent time in a group, and have found it is very significant \citep[e.g][]{McGee2009,DeLucia12}. For example, \cite{Han2018}, using hydrodynamic zoom-in simulations, recently found that around 48\% of today's cluster members were previously members of other host halos. Moreover, their results consider hosts of all masses to better constrain the effect of a previous environment. Furthermore, this study suggests that on average half of the dark matter mass loss of cluster galaxies happened prior to infall into the cluster. From an observational point of view, several studies have indeed revealed substructures within galaxy clusters \citep[e.g][]{Aguerri2010,Jaffe2016,Jauzac2016,Tempel2017,Lisker2018}, which highlights their complex, hierarchical assembly. There is a growing body of evidence that the group environment alone can influence its satellites. The low internal velocity dispersion of groups may greatly enhance galaxy-galaxy interactions \citep[e.g.][]{Paudel2014,Ooterloo2018}. There is also evidence that star-formation quenching of satellites occurs in groups before they infall into clusters \citep{Rasmussen2012,Bianconi2018,Smith2019}. In fact, the physical processes suppressing star formation, such as ram-pressure stripping, may also contribute to the removal of cold galactic gas in group galaxies \citep{Rasmussen2006,Brown2017}.

The cluster environment itself may result in highly efficient tidal stripping from gravitational interactions between galaxies and interactions with the cluster potential \citep{Smith2010,Bialas15,Smith2015}. Together with the effect of the intra-cluster medium these interactions can alter the structure of a galaxy, but the degree to which it is affected also depends on whether is was previously pre-processed. For example, dark matter halos that were previously in a group environment tend to lose mass more slowly in the cluster environment as the more weakly bound dark matter in the outer halo was already removed in the group \citep{Joshi2017,Han2018}. Also, in some Virgo cluster dwarf ellipticals, counter-rotating cores have been observed \citep{Toloba2014}. Such features are very difficult to reproduce through the action of the cluster tides or harassment alone \citep{Gonzalez-Garcia2005}. Thus mergers in a previous group environment are likely required to explain the presence of such features in galaxies that are now cluster members.

 In addition to pre-processing, it is important to take into account that galaxies entering the cluster in a group face the combined environmental influence of both group and cluster, and it may be several gigayears until groups can be completely dissociated and virialized within the cluster \citep{Cohn12}. This process, arising in group-cluster mergers, is often referred to as \textit{post-processing}. In \cite{Vijayaraghavan2013} this process was studied using N-body cosmological simulations combined with idealized N-body/hydrodynamical simulations. They study the effects that the merger has on both group and cluster: there is an enhancement of the galaxy-galaxy merger rate during the pericentric passage of the group in the cluster; the merger shock leads to an increase in the ram pressure acting on both the group and the cluster galaxies; and the group galaxies may become tidally truncated.

In this context it is clear that to fully understand the processes that are triggering changes in the cluster galaxy population it is crucial to understand their environmental history. While the quantity of pre-processed galaxies in clusters has been shown to be quite significant, as has the effect of the group environment upon them, our understanding of what happens to these galaxies once their hosts enter the cluster, the post-processing stage, has not been explored in so much detail.

Thus, in the study presented here, we focus on the consequences of the group falling into the cluster environment. Using a set of high resolution N-body simulations of galaxy clusters, we identify host and satellite halos at different redshifts, and then track their evolution while they are under the influence of the cluster environment. We specifically include satellites that are stripped from their host halo by the cluster tides in order to better understand the role of the cluster in the dissociation of substructure.  Understanding how groups break up in clusters could potentially help us identify these pre-processed galaxies, once they have left their groups and mixed with the cluster population. However, as we will show later in this study (Section \ref{sec:results}), we find that the manner in which satellites leave their hosts makes this quite challenging.

The paper is organised as follows. In Section \ref{sec:method} we describe the cosmological zoom-in simulations of clusters and the satellite-host membership criterion we used. The evolution of the satellite sample is presented in Section \ref{sec:results}. Finally, we discuss and summarise our key results in Section \ref{sec:discussion} and \ref{sec:conclusions} respectively.

\section{Method}
\label{sec:method}

\subsection{Simulations}

To carry out this analysis we use N-body simulations performed originally using the Tree-code GADGET 2 \citep{Springel2001,Springel2005}. These simulations used a $\Lambda$CDM (cold dark matter) cosmology with initial conditions of $\Omega_{m,0}$ = 0.3, $\Omega_{\Lambda,0}$ = 0.7, $H_0$ = 68 km/s/Mpc, $\sigma_8$ = 0.82 and $n=0.96$ in a 140 Mpc/h box that was evolved from $z=200$ to $z=0$, with particle masses of $1.7\times 10^9$ M$_\odot$/h and with a softening length of 5.469 kpc/h. The initial conditions were generated using MUSIC \citep{Hahn2011} and CAMB \citep{Lewis2002}. At $z=0$ the halos were identified by using a parallel friends-of-friends algorithm \citep[FoF;][]{Kim2006} and a subset was re-simulated with a higher mass resolution of $3.32\times10^6$ M$_\odot$/h per particle and with a reduced softening length of 0.683 kpc/h. The size of the zoom-in region encompasses all particles within 5.5R$_{200}$ of the cluster at $z=0$ following the prescription of \citet{Onorbe2014}.

Given the large size of the zoom-in region, we expect little contamination from low resolution dark matter particles within the clusters. We find only a small number of low resolution dark matter particles (53) within 1R$_{200}$ of the cluster, which is only $\sim2.1\times10^{-6}$ of the total in that radius.

The halos of the zoom-in simulations and their properties were measured using the Amiga Halo Finder (AHF; \citealp{Knollmann2009}). The merger tree used to connect progenitor/descendant halos was constructed using our own routine in which we track the most bound particles.  These simulations have sufficient resolution to enable us to follow the formation and evolution of subhalos down to a minimum halo mass limit of $\sim6.64\times10^7$ M$_\odot$/h. This is set by the minimum
number of particles necessary to define a halo, which is chosen to be 20. We use 120 output snapshots from $z=9$ to $z=0$. Two main cluster halos at $z=0$ with masses of $\sim10^{14}$ M$_\odot$/h are used in this analysis as is shown in Table \ref{tab:table1}. Further details of the simulation can be found in \citet{Taylor2019}.

 In this analysis we use ``cluster" halo to refer to the most massive halo in each re-simulation while any other halo hosting at least one less massive halo (``satellite" henceforth) is referred as a ``host" halo. We explain the satellite membership criterion in the next section (\ref{sec:membership criterion}). Furthermore, throughout this study the spatial extent of our halos is defined in terms of R$_{200}$, which corresponds to the radius within which the mean density is 200 times the critical density of the universe. In the same way, M$_{200}$ is defined as the total mass enclosed within that radius.

\subsection{Membership Criterion}
\label{sec:membership criterion} 
Before applying any criterion on the full sample of halos, we discard all those below a chosen limit on their peak masses. We set this limit at M = 2.0$\times 10^{8}$ M$_\odot$/h, i.e, we only consider halos with a peak mass greater than this value. In this way, the effects of the resolution are reduced, and we are able to follow the mass loss of even the lowest mass halos down to at least 30$\%$ before they reach the resolution limit. 

\begin{figure}
	\includegraphics[width=\columnwidth]{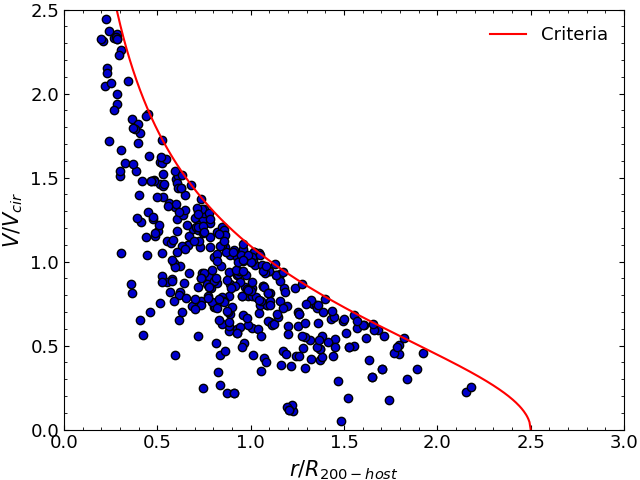}
    \caption{Phase-space diagram for the satellites in the host-centric frame at the cluster infall time. The x-axis is the separation distance of the satellite with respect to its host normalised by the R$_{200}$ radius of each host, while the y-axis is the relative velocity normalised by the circular velocity at R$_{200}$ of the host. The red line highlights the satellite membership criterion used. Note that the data of both clusters are plotted here. }
    \label{fig:fig1}
\end{figure}

Our main focus is to track the evolution of those halos that are bound to more massive host halos that are subsequently accreted into the cluster. Therefore, we use a similar criterion as in \cite{Han2018} to define satellites and host halos in the simulations. As explained in that study, the use of the R$_{200}$ radius as a limiting radius to classify a halo as bound to another is too restrictive as we may mistakenly classify some satellites (e.g ``flyby" galaxies) or fail to classify others. For example, by definition backsplash galaxies \citep{Balogh2000} will exit the host R$_{200}$ radius before falling back in. Therefore, the criterion used extends the maximum radius out to 2.5R$_{200}$ to ensure these galaxies are not missed as group members. As can be seen in the lower right corner of Figure \ref{fig:fig1} (also Figure 1 in \citealp{Han2018}) there are few points in that region, thus 2.5R$_{200}$ is in general sufficient. The criterion is defined as follows:

\begin{equation}
   \centering
    \frac{v^2}{2} +\Phi(r) < \Phi(2.5R_{200}).
	\label{eq:equation1}
\end{equation}

The right hand side of the inequality (\ref{eq:equation1}) is the host gravitational potential $\Phi(r)$ at 2.5R$_{200}$ from the centre of the host, while the left hand side is the specific orbital energy of the satellite with respect to its host, $(v)$ being the relative velocity between them. When a subhalo satisfies this inequality we can identify it as a satellite of its host halo. We apply this binding criterion at the moment that hosts fall into the cluster (i.e. when they cross R$_{200}$), meaning they are defined at a variety of redshifts. We cannot define them earlier, or at a fixed time, as hosts continuously accrete new members outside of the cluster. We will study when the accretion of new satellites stops within the cluster in Section \ref{sec:Satellite accretion...}. Finally we have 146 hosts and 394 satellites identified in the two clusters as is shown in Figure \ref{fig:fig1} (and Table \ref{tab:table1}), where the position of every single identified satellite is highlighted on the phase-space diagram in their host-centric frame. In this figure, the distance (normalised by R$_{200}$) is plotted against the relative velocity (normalised by the circular velocity at R$_{200}$ of the host). The region below the line highlights the bound region according to the criterion used.

The distribution of the satellites in the phase-space diagram at the time of crossing the cluster R$_{200}$ radius (defined hereafter as the cluster infall time) is broad, with a few of them being in the ``back-splash" region (bottom right), while others are located in the region close to first pericentric passage (upper left), and some populate the region of ``ancient infallers'' (bottom left), which are approximately virialised with the host potential well. We will later study the differences in the satellite distribution within their hosts in more detail.

\begin{table*}

 \label{table_example}
  \begin{tabular}{l||c|c|c|c|r}
   \hline
&M$_{200}$ (M$_\odot$/h)& R$_{200}$ (kpc/h)&$z_{form}$& $N^o$ of satellites& $N^o$ of hosts\\
     \hline\hline
cluster1 &$9.16\times 10^{13}$& 733.22 &9 &187& 71\\
      \hline
cluster2 &$1.04\times 10^{14}$& 765.19 &7.5  &207& 75\\
      \hline
   \end{tabular}

    \caption{Properties of the two clusters at $z=0$, where M$_{200}$ is the mass, R$_{200}$ is the outer radius and $z _{form}$ is the redshift of formation. The last two columns correspond to the number of satellite and host halos in each cluster.}
    \label{tab:table1}
\end{table*}

\subsection{General properties of satellites and their hosts at infall time}

The distribution of the number of satellites in a host is broad, with most of the hosts (around 63 \%) in our sample having only one member, while around 34\% host between 2 and 17 satellite members and $\sim$ 3\% host more than twenty, these being the most massive hosts in our sample. In addition, we have a broad host mass range of $\sim 3\times 10^8 - 1\times 10^{12}$ M$_\odot$/h (dwarfs to roughly Local Group-mass halos) while their R$_{200}$ radii are in a range of $\sim 13-183$ kpc. Note that there are no ongoing major mergers taking place in the two clusters. The satellite masses are in the range $1\times10^{8}-5\times 10^{10}$ M$_\odot$/h.  We note that there are four low mass host halos ($ < 1 \times 10^9$ M$_\odot$/h). This is possible because we do not put any constraint on the mass of the host: it is simply required to be more massive than its satellite, and the satellite must satisfy the binding criteria given in the inequality in (\ref{eq:equation1}).

\subsection{Classification of outcomes for the satellites at $z=0$}
\label{sec:outcome of sat...}

\begin{figure}
	\includegraphics[width=\columnwidth]{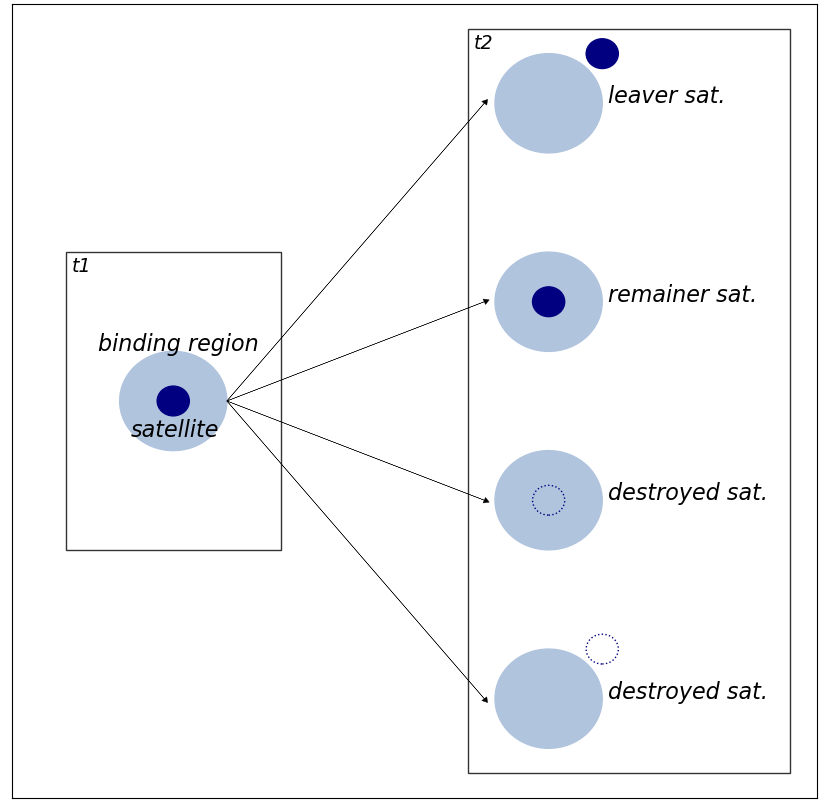}
    \caption{Schematic figure showing the three possible outcomes for satellites. Shown in the left rectangle is the initial time (t1) when the satellite is within the binding region of its host halo, while the right rectangle shows the possible outcomes at some later time (t2) where the satellite has left the host, has remained within it or has become ``destroyed'' considering the two possible options for this (either inside or outside the binding region of their host).  }
    \label{fig:fig2}
\end{figure}

We track the evolution of the hosts and their satellites once they enter the cluster until redshift $z=0$. By using the binding criterion explained in Section \ref{sec:membership criterion} we are able to identify whether a satellite still remains bound by the end of the simulation. Thus we identify the possible outcomes as shown in the schematic in Figure \ref{fig:fig2} which we describe and define as follows.

\begin{enumerate}[I]
  \item \textit{Leaver satellite}: At some point during their evolution in the cluster these satellites become unbound from their host. At $z=0$ it is no longer a host member. Note that through this study we use both terms, ``leave" and ``unbound" interchangeably. 
  
  \item \textit{Remainer satellites}: Contrary to leavers, these objects are still bound to their hosts at redshift zero. 
  
  \item \textit{Destroyed satellites}: Because of tidal mass loss, some halos may fall below our mass resolution limit. This can happen when a satellite is still bound to its host or once it is already unbound (as is shown in the last two schematics, lower-right corner of Fig. \ref{fig:fig2}  ). As long as the halo is resolved we can track the tidal mass loss, but once it disappears below the resolution limit we cannot differentiate between tidal mass loss and a merger. Note that some of the hosts of these types of satellites may also be ``destroyed" at some point.
\end{enumerate}

Under this classification we identified 128, 48, and 213\footnote{Note there are 5 satellites that do not belong to any classification. This is because these satellites become unbound first but later their hosts are destroyed. In principle they could fit in the category of ``leaver'', however our definition of ``leaver'' is such that the hosts are not destroyed at $z=0$.}, leavers, remainers and destroyed respectively, combining the data of the two clusters. In general, as can be seen, the destroyed population is the largest. We will investigate this population later (section \ref{sec:delving more...}). For now, in this study we wish to understand why and when a satellite becomes unbound and leaves the host. We explore this in the next section.

\section{Results}
\label{sec:results}

\subsection{When do satellites leave the host halo?}
\label{sec:when do sate...}

Before addressing this question we first describe the orbits of the hosts in the clusters. As some hosts may have only recently infallen, they might not have had a pericentric passage by $z=0$. 89.7\% of our hosts had at least one pericentric passage while only 10.3\% have yet to reach pericentre. For example, almost all of the satellites (92.2\%) in groups that pass pericentre leave their groups either before of after the passage, while the rest, which corresponds to only $\sim$ 7.8\%, remain bound to their host until $z=0$, despite having had a pericentric passage.

\begin{figure}
	\includegraphics[width=\columnwidth]{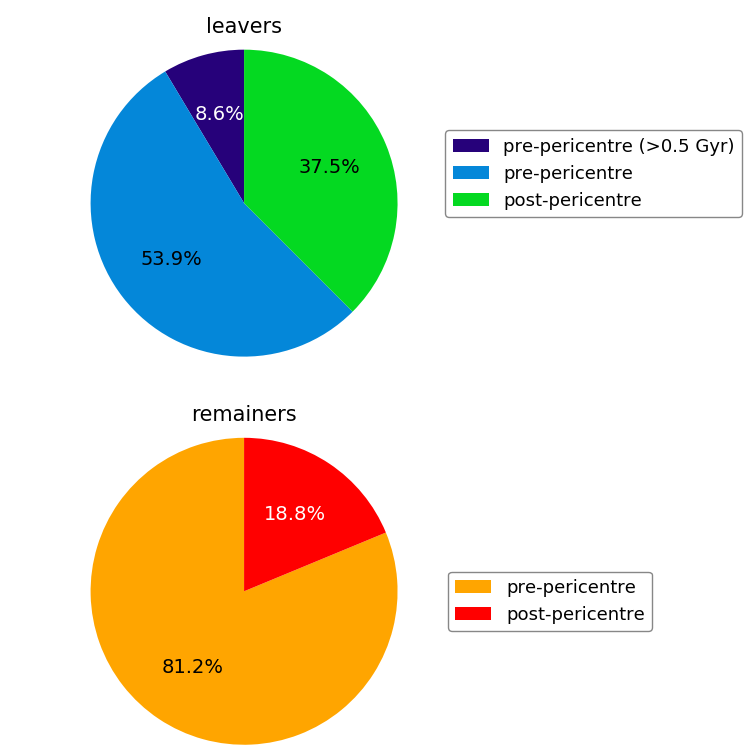}
    \caption{Top: Percentages for leaver satellites classified as pre-pericentre (blue), early leavers which leave earlier than 0.5 Gyr before pericentric passage (dark blue), and for those classified as post-pericentre (green). Bottom: Percentages for remainer satellites classified as pre-pericentre (orange) and as post-pericentre (red). Note that only survivor satellites are considered here (i.e. no destroyed).}
    \label{fig:fig3}
\end{figure}

We now split the satellites according to whether they are leavers or remainers, and whether they have completed their first pericentric passage of the cluster yet. The fractions of each are shown in the piecharts in Figure \ref{fig:fig3}. We first consider the top piechart. Around 62.5\% leave the host before the passage (labelled as pre-pericentre). Note that in this category we also include those satellites that leave from a host that does not reach pericentre by $z=0$. Only a few of the leaver satellites ($\sim$ 8.3\%) leave early-on,  (earlier than 0.5 Gyr before pericentric passage, labelled as pre-pericentre (> 0.5 Gyr) and referred to as `early leavers'). Thus most of the pre-pericentre leavers leave only shortly before pericentre (we examine the time at which they leave further in Figure \ref{fig:fig5}). On the other hand, the other 37.5 \% leave their hosts after the pericentric passage (labelled as post-pericentre). In this last case, we tested when they leave and find that most of these ($\sim$ 80\%) leave quite shortly (within 0.5~Gyr) after the first pericentric passage.  Now we consider the bottom piechart. 82.2\% of the remainer satellites are in hosts that have only recently fallen into the cluster and so have yet to reach their first pericentre by $z=0$ (labelled as remainer pre-pericentre). The others, which corresponds to $\sim$ 18.8\%, remain bound to their host until $z=0$ despite having had a pericentric passage.

\begin{figure*}
	\includegraphics[width=\textwidth]{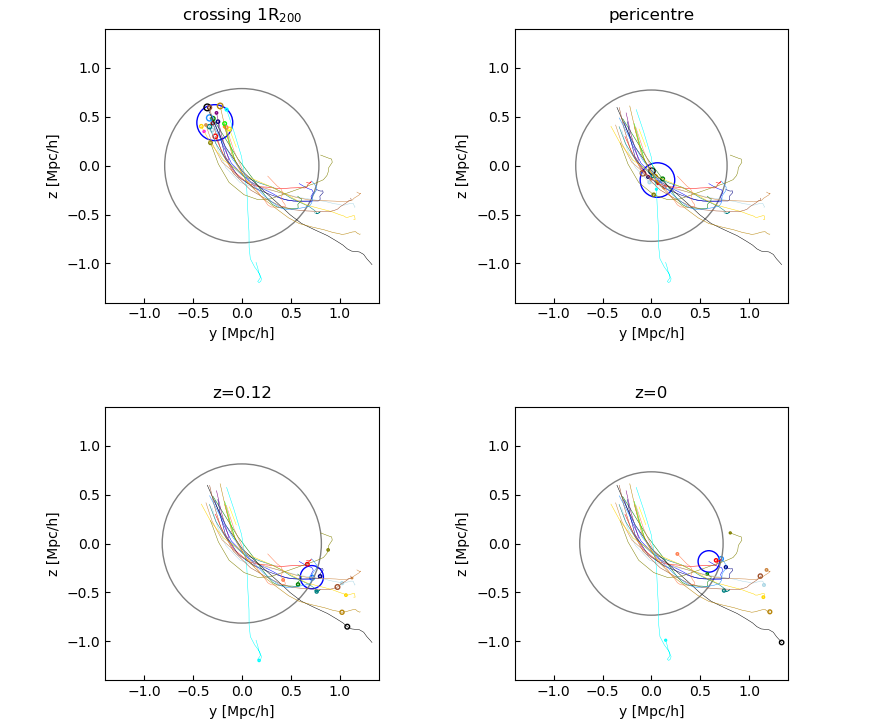}
    \caption{Schematic projected figure showing the break up of one of the host halos in a time sequence; crossing 1R$_{200}$ in the cluster, at pericentre time, some time later, and at redshift zero. The grey circle represents the cluster, the blue circle represents the host and the other coloured circles represent its satellites. Lines with the same colours show their trajectories through the cluster for all times after crossing the R$_{200}$ radius. Note that the choice of projection
means the group crossing R$_{200}$ appears slightly inside the R$_{200}$ circle of the cluster. Also, by $z=0$ there are less satellites as some of them become classified as destroyed by this time.}
    \label{fig:fig4}
\end{figure*}

\begin{figure}
	\includegraphics[width=\columnwidth]{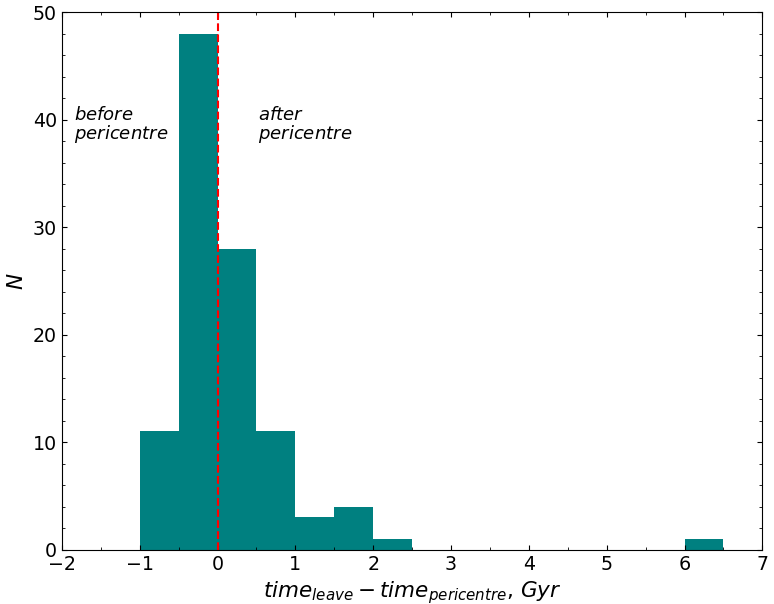}
    \caption{Distribution time at which a satellite becomes unbound from its host with respect to the first pericentric passage of the host. The bars to the left of the vertical red line correspond to satellites leaving before pericentre, while the bars to the right correspond to satellites leaving during or after this moment.}
    \label{fig:fig5}
\end{figure}

These results clearly highlight how most of the substructure has been broken down by the time it has passed pericentre just once. In Figure \ref{fig:fig4}, we provide a visual example of the dissociation of a group when it passes cluster pericentre for the first time, shown as a time sequence. The projected positions of the cluster, host and satellite halos are shown in each panel as circles. The circle radii are equal to the R$_{200}$ radii of the halos. The first panel shows the moment we define the satellites of that group, when the group first crosses R$_{200}$ of the cluster. At this time, all (20) satellites are bound to the host. The second panel shows the moment when the first pericentric passage occurs. Already, some of the satellites have been destroyed, while the other halos have been unbound and start to move away from the host halo. The third panel shows the moment when the halo reaches first apocentre. By this time, one more satellite has been destroyed and the other satellites are now far from their previous host. Finally, the last panel shows the positions at $z=0$. The host is now approaching the cluster once again, and some of the satellites have turned around as well. But the satellites are all unbound and far away from their former host, and also distant from each other.

In connection with this in Figure \ref{fig:fig5} we show when (exactly) a satellite leaves its host. The x-axis corresponds to the time when a halo leaves relative to the time of pericentric passage of its host, with negative values being before pericentre, and positive values being after. From the Figure \ref{fig:fig5} we note that this often happens shortly before the pericentric passage, usually within 0.5 Gyr (see the histogram bars to the left of the vertical line). As in the piecharts in Figure \ref{fig:fig3}, we refer to these objects as pre-pericentre leavers. They make up roughly $\sim$ 54\% of the total population of leavers. This means that most of the unbinding happens before the tidal shock within the cluster centre. Thus, it appears that the cluster tides near pericentre, alone, are sufficient to separate the hosts from their satellites. The others leave their hosts either at the moment of pericentric passage (i.e., the histogram bar at 0 in the x-axis) or shortly afterwards. Of these, most (80$\%$) leave within 2 Gyr after pericentre, but in a few rare cases they leave later. For these very late leavers, we checked if this occurs at a later pericentric passage. However, most of these satellites become unbound before they reach second pericentric passage, perhaps because of steady additional mass loss from the cluster tides, or because of an unusually long delay in satellites separating from their hosts.

\begin{figure}
	\includegraphics[width=\columnwidth]{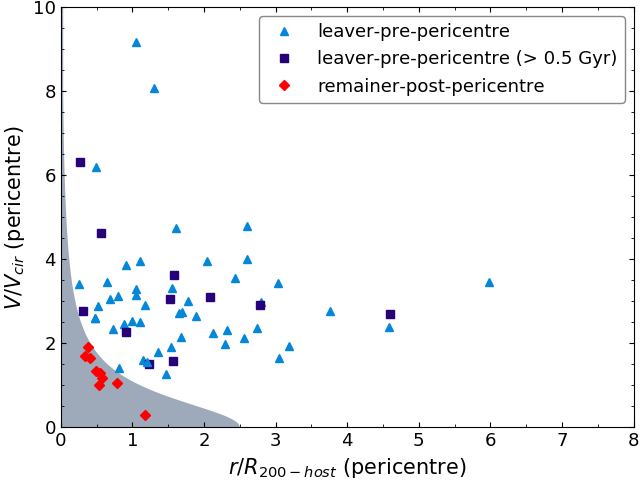}
    \caption{Phase-space diagram for satellites in the hostcentric frame at the instant of first pericentre. Leaver satellites classified as pre-pericentre are indicated with blue triangles, early leavers which leave 0.5 Gyr or more prior to first pericentric passage are indicated with dark blue squares, and those classified as remainers post-pericentre are indicated with red diamonds. The grey region highlights the binding region. Note that only the satellites in hosts having a pericentric passage by $z=0$ in the cluster are included here.}
    \label{fig:fig6}
\end{figure}

\begin{figure}
	\includegraphics[width=\columnwidth]{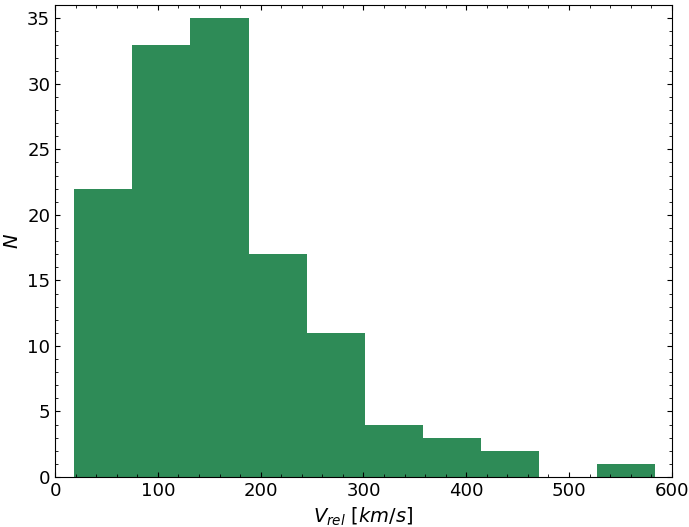}
    \caption{Distribution of relative velocities with respect to their hosts of the leaver satellites at the moment they become unbound.}
    \label{fig:fig7}
\end{figure}

In addition, we also checked where the satellites are located in their hosts when they are at the first pericentric passage in the cluster. Figure \ref{fig:fig6} shows the phase-space diagram in their hostcentric frame at this time. The colour scheme follows that established in Figure \ref{fig:fig3}, with pre-pericentre leavers shown in blue (triangle symbols) and early leavers shown in dark blue (square symbols). For completeness we also include the remainer (post-pericentre) satellites that, despite passing the cluster core, remain bound (red diamond symbols). We see that these satellites are closer to their host centres, within the grey shaded region that represents the binding region. Meanwhile the leavers are widely distributed in both distance and velocity from their hosts. Given that Figure \ref{fig:fig5} tells us that most pre-pericentre leavers exit their host only shortly before pericentric passage, this shows that the unbound satellites must quickly separate from their hosts. We measure the relative velocity that the leaver satellites have at the moment they become unbound from their hosts as shown in Figure \ref{fig:fig7}. It can be seen that they leave with a wide range of velocities, from $\sim$10 km/s up to an extreme $\sim$580 km/s. However, even with a more modest central value of 150 km/s, it can easily be shown that satellites will move a considerable distance with respect to their host over a short time period, as seen in Figure \ref{fig:fig6}. Therefore, this implies that it will quickly become difficult to identify them within the cluster population at later times. We analyse the causes and implications that this result has for the satellites and their hosts further in section \ref{sec: causes: separation...}.

\subsection{Why do the satellites become unbound?}
\label{sec: causes: separation...}

In this section we investigate the causes for satellites becoming unbound from their hosts. We consider three parameters: initial separation, host tidal mass loss and separation at the host's pericentre time.

\begin{figure}
	\includegraphics[width=\columnwidth]{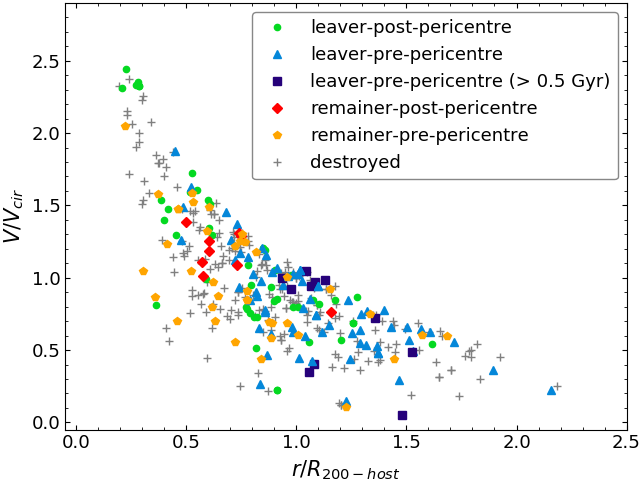}
    \caption{Phase-space for satellites in the hostcentric frame at the cluster infall time (time when we measure our binding criterion). The colours and symbols highlight the different subsamples of satellites following the same colour scheme as in Fig. \ref{fig:fig6} with the addition of several other subsamples; leaver post-pericentre (green dots), remainer pre-pericentre (orange pentagon symbols), and grey crosses correspond to satellites that are destroyed in the cluster.}
    \label{fig:fig8}
\end{figure}

In Figure \ref{fig:fig8} we show the phase-space diagram for satellites in the hostcentric frame at the cluster infall time (when we measure our binding criterion; same as Figure \ref{fig:fig1}). The idea is to understand how the outcome for every satellite at $z=0$ is related to their location in the group at infall. For completeness the ``destroyed" halos are also included here. Also, we split satellites into categories, following the same colour scheme as in Figure \ref{fig:fig3}. Overall, the post-pericentre leavers are distributed throughout phase-space. This indicates that, if they pass pericentre, they will leave regardless of where in the group they were located at infall. However, the pre-pericentre leavers (blue triangle and dark blue square symbols) are more likely found in the outer regions of their host at the time of infall, where they are more weakly bound. This is especially true for the early leavers (dark blue squares) which are mostly found beyond one R$_{200}$ radius of their host at cluster infall time. This is in contrast to the red points (satellites that remain bound to their host after pericentre), which are generally restricted to within one R$_{200}$ radius from their hosts. Pre-pericentre remainers (orange pentagon symbols) must be bound to groups on first infall into the cluster, and are distributed quite evenly across the diagram. This is also the case with destroyed halos (grey crosses), indicating that location within the group at the time of cluster infall does not play a strong role in deciding whether a halo will be destroyed. This highlights that host tides do not play an important role in destroying halos, a point we will see further evidence for in Section \ref{sec:delving more...}.

\begin{figure}
	\includegraphics[width=\columnwidth]{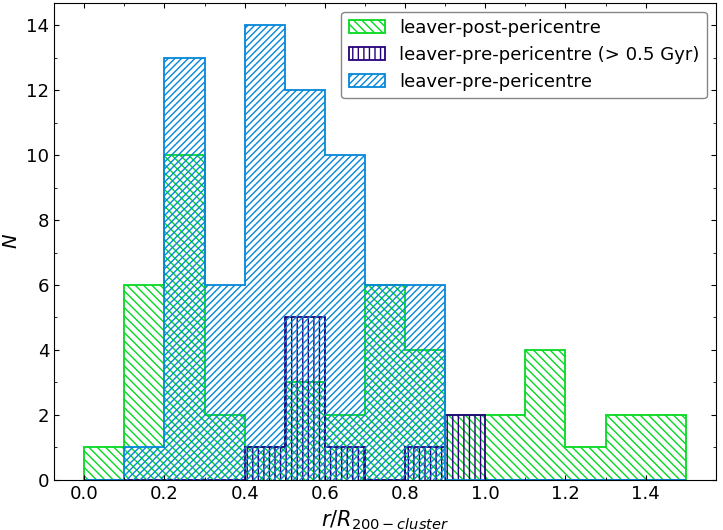}
    \caption{Distribution of the clustercentric distances of leaver satellites normalised by the cluster R$_{200}$ radius at the moment they become unbound from their hosts. Colours in the histograms highlight the three subsamples for leavers, as in Fig \ref{fig:fig3}.} 
    \label{fig:fig9}
\end{figure}

We show the clustercentric radius (normalised by the cluster R$_{200}$ radius) at which satellites leave their hosts in Figure \ref{fig:fig9}. For those leaving before the pericentric passage, the distribution is wide but peaks at about 0.45 R$_{200}$ (blue bars). The early leavers have a narrower distribution which peaks a little further out, indicating they leave in the outer half of the cluster, shortly after infall in to the cluster. For those leaving after the pericentric passage (green bars), there is a broader distribution, extending beyond the cluster R$_{200}$ radius. This indicates that they leave over a wide range of distances, and can leave even when their host has backsplashed (i.e. has passed pericenter and is found beyond the cluster R$_{200}$ radius at the time of leaving).

\begin{figure*}
	\includegraphics[width=\textwidth]{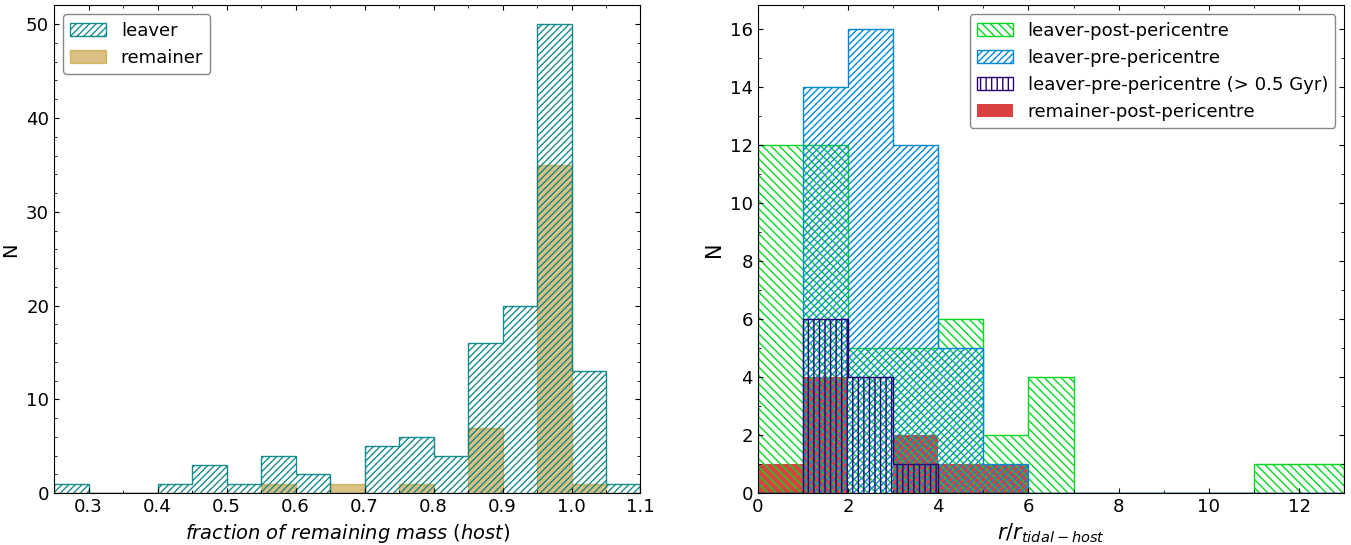}
    \caption{Left panel: Distributions of the fraction of remaining mass for host halos. Teal bars correspond to the leavers' hosts while pink bars corresponds to the remainers' hosts. Right panel: Distribution of hostcentric distances of satellites normalised by the host tidal radius, at the moment when the host first passes cluster pericentre for remainers and at the moment satellites become unbound for the leavers. As usual, the colours of the histograms correspond to the same colour scheme as in Fig. \ref{fig:fig3}, however note that only satellites in hosts that have passed pericentre by $z=0$ are included here.}
    \label{fig:fig10}
\end{figure*}

We might expect that the tidal mass loss of hosts is closely related with the outcome of their members, i.e. that hosts losing a high fraction of their mass might more easily release their member satellites. Thus, in the left panel of Figure \ref{fig:fig10} we show the distribution of the fraction of remaining mass for the hosts. For the hosts of remainer satellites this fraction is $m_{z=0}$/$m_{infl}$, the mass at $z=0$ divided by the mass at the cluster infall time. For those hosts of leaver satellites this fraction is $m_{leav}/m_{infl}$, mass at the moment the satellite leaves divided by the mass at cluster infall time. The distribution of the remaining mass for the leaver's hosts is wide (teal bars), with some of them losing more than 50\% of their masses while others lose a smaller percentage ($< 20\%$) yet still have leaver satellites. In a few cases some hosts even gain a small amount of mass. In the case of the hosts of remainers (yellow bars), they have overall lost a smaller percentage of their mass, but this is related to the fact that in most of these cases they have not yet reached cluster pericentre, and so they would presumably lose more mass. Therefore, it appears that the probability of a host having a leaver satellite is not a sensitive function of the amount of host mass loss.

We can gain more insight by considering the tidal radius of the hosts, which is strongly correlated to the radial location of the host in the cluster. In the right panel of Figure \ref{fig:fig10} we show the distribution of the hostcentric distances of the satellites, normalised by their host's tidal radius ($r_{tidal}$), at the time satellites become unbound for the leavers, and at the time of pericentre for the remainers, when the cluster tides are strongest. The tidal radius is calculated in a simplistic way by assuming a constant background tidal field from the cluster, $r_{tidal}=r_{ho-cl}(m_{ho}/2M_{cl}(<r)) ^{1/3}$, where $r_{ho-cl}$ is the host-cluster separation, $m_{ho}$ is the host mass, and $M_{cl}(<r)$ is the cluster mass enclosed within the host's clustercentric radius. From the figure it can be seen that remainer satellites tend to be at smaller distances from their host, compared with the leaver distributions (blue, green and dark blue) which tend to have tails at much larger separations from their host (with the one exception of the early leavers). Note that the red histogram is an upper limit on the typical $r/r_{tidal-host}$ for the remainers, because in that case we use the pericentric distance.
The larger separations indicate that satellites close to their host centre at the moment of pericentric passage are much less likely to leave than those distributed further out. This makes physical sense when one considers that the differential acceleration from the cluster potential, which acts to split the satellite from its host, is a sensitive function of the satellite-host separation. For example, even with extreme background tides from the cluster, causing strong tidal mass loss, if the group member is at the very centre of the group then the differential acceleration is zero, and they will not be separated by the tides. We note that leaver satellites that become unbound before the pericentre are also included in this figure.

\subsection{Tidal mass loss of hosts and their satellites}

\begin{figure*}
	\includegraphics[width=\textwidth]{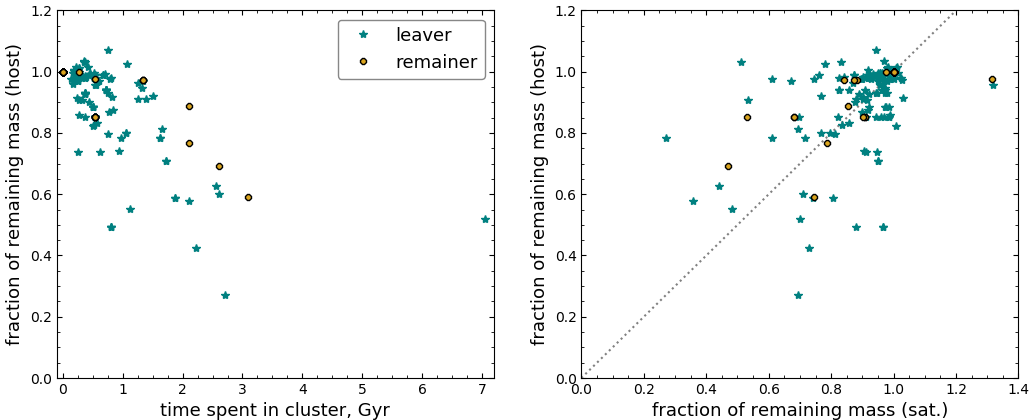}
    \caption{Left panel: Fraction of remaining mass for the host halos as a function of the time that they have spent in the cluster. This time depends on whether they are the host of remainer (yellow dots) or leaver (teal star) satellites. Right panel: Fraction of remaining mass for satellites versus hosts. The line indicates a 1:1 relation between the two parameters. }
    \label{fig:fig11}
\end{figure*}

In addition to the previous results, in the left panel of Figure \ref{fig:fig11} we show the fraction of remaining mass for our host sample as a function of the time that they have spent in the cluster. As before, this time is measured since the cluster infall time until $z=0$ (for the remainers' hosts; yellow dots) or at the time the satellites become unbound (for the leavers' hosts; teal star symbols). This is because we are interested only in mass loss occurring while within the host.  Overall, it can be seen that the fraction of remaining mass for hosts decreases, i.e. there is more mass loss as the time spent in the cluster increases. For example, those that have spent more than 2 Gyr have lost $> 40\%$ of their mass. This means that hosts entering the cluster earlier lose more mass, because they have more time to be affected by the cluster tidal forces, in comparison to those entering later. We also may take into account the fact that the hosts of the leaver satellites eventually will lose more mass, as their mass is taken at a higher redshift and not at $z=0$ as for the hosts of the remainers. Thus these hosts will presumably continue to evolve towards the bottom right region of the figure. We do not see a significant difference between the mass loss of remainer and leaver hosts.

In the right  panel of Figure \ref{fig:fig11}, we plot the efficiency of mass loss in satellites and the relation with the mass loss of their hosts. Overall, we can identify three regions here. Points situated above the 1:1 line show satellites that have lost more mass than their hosts. This could occur due to the satellites suffering the combined tides of the host plus that of the cluster, meanwhile the host only suffers from the cluster tides. We should also take into account that satellites already start losing mass in their hosts before they enter the cluster (i.e. pre-processing). We also see points close to the 1:1 line, which indicates satellites that are more or less as equally affected as their hosts. On the other hand, some points are situated below the 1:1 line, indicating that their hosts lose a larger fraction of mass than their satellites. We checked the reason for this by measuring their distances to the cluster centre as well as the clustercentric distances of their hosts at the (first) pericentre time or at $z=0$ for those hosts still on first infall by that time. We found that in the majority of these cases, the hosts are closer to the cluster centre than their satellites, and so they are more exposed to the cluster tides.

For a subsample of satellites that suffer significant tidal mass loss (defined as losing $>10\%$ of their dark matter inside the cluster), we calculate the percentage that lose more mass than their host (i.e. fall above the 1:1 line). For the remainers and leavers combined this is $\sim76.3\%$, meaning most of the time satellites lose more mass than their host, as a result of the additional tides they suffer from the group tides. Separately, the percentage is $\sim81.8\%$ for the remainers and $\sim75\%$ for the leavers. This means the remainer satellites suffer more mass loss from their host than the leavers, likely because they tend to spend more time in the presence of their host. 

These results show that, generally, the presence of the group causes a weak but not very significant increase in the tidal mass loss of the satellite within the cluster. In fact, the group may have caused much larger tidal mass loss to its satellite outside the cluster. But we are only measuring the tidal mass loss occurring within the cluster (i.e. post-processing) and this is weak because there is limited time before the substructure is disrupted shortly after first pericentre.

\subsection{Where do the satellites end up? }

\subsubsection{Final location with respect to the host halo}

Further to our analysis of the evolution of the satellites and their hosts when they become part of the cluster, we now show where the satellites are at $z=0$ in a more detailed way than already described in Section \ref{sec:outcome of sat...}. Thus, in Figure \ref{fig:fig12} (left panel) we show first the phase-space diagram in the hostcentric frame. Remainer satellites, as expected (and by definition) are in the bound region and, as we already discussed, a few of them ($\sim$2.3\% with respect to the full sample) are bound to the host despite the host having passed close to the cluster centre (red diamond symbols; classified as post-pericentre remainers). We note that they typically have small separations from their hosts at $z=0$ ($<$ 1R$_{200}$). So those few hosts whose satellites remain bound after first pericentre are much more compact afterwards. {\it{This implies that any extended galaxy groups that are observed in clusters are likely to be on first infall.}} Regarding the others (orange pentagon symbols; classified as pre-pericentre remainers), they still remain bound within hosts that have not yet had a pericentric passage but are still on first approach by $z=0$. It can be seen, however, that most of the leaver satellites are very far from their original host (for both the pre- and post-pericentre subsamples; green, dark blue and blue symbols). Some are at distances of greater than 20 times their host's R$_{200}$ radius and with very high velocities. Those leavers (blue triangles) that are close to the bound region are those that became unbound very recently, as their hosts are just approaching the cluster centre. Overall, these results indicate that many post-processed galaxies are well mixed in with the rest of the cluster population today, and very difficult to associate with their original hosts.

\subsubsection{Final location with respect to the cluster}

\begin{figure*}
	\includegraphics[width=\textwidth]{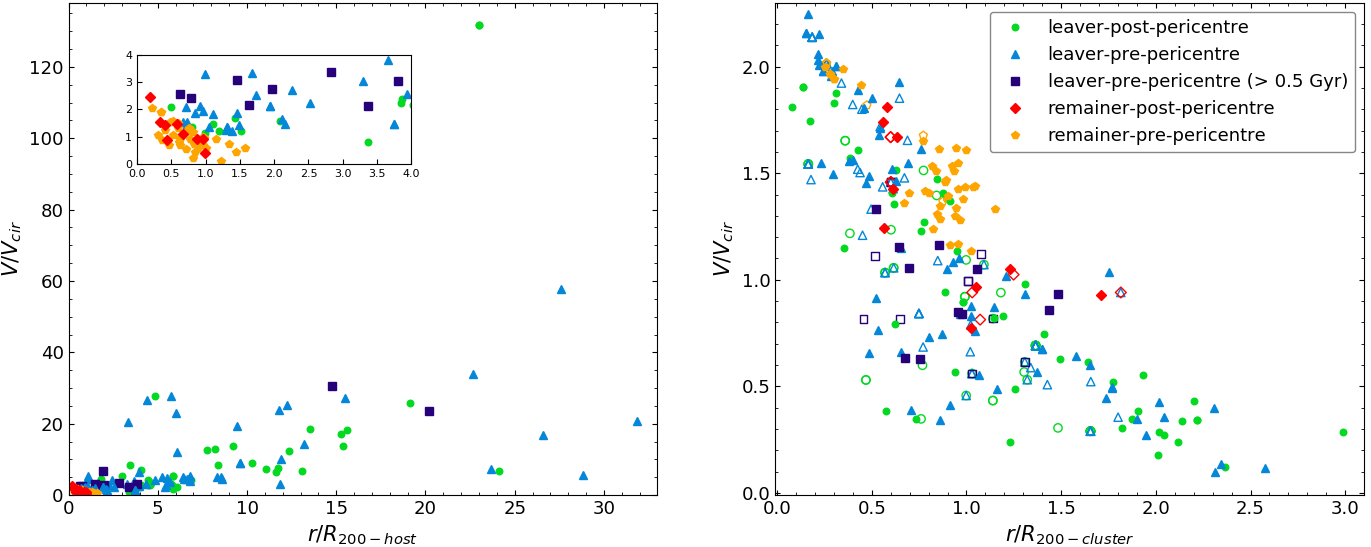}
    \caption{Left panel: Phase-space diagram for satellites in the hostcentric frame at $z=0$. The subfigure shows a zoomed region of the same plot. Right panel: Phase-space diagram for satellites (filled symbols) and hosts (open symbols) in the clustercentric frame at $z=0$. The colours and the different symbols in both panels indicate the five subsamples of satellites following the same colour scheme as in Fig. \ref{fig:fig3} and Fig.\ref{fig:fig8}. }
    \label{fig:fig12}
\end{figure*}

In Figure \ref{fig:fig12} (right panel), we show the phase-space diagram in the clustercentric frame at $z=0$. As before, we include the four subsamples for satellites (filled symbols) as well as their hosts (open symbols). Overall, points can be seen throughout the different regions of the phase-space diagram, with some of them found in the backsplash region beyond one R$_{200}$ radius, and with low velocity (bottom right), while others are found in the ``ancient infaller'' region (small distance, low velocity; bottom left), and some are found in the near first-pericentre region (small distance, high velocity; upper left). 

 It can further be seen that leaver satellites are widely distributed and have the largest distances to the cluster centre ($>2$R$_{200}$). Interestingly, we note that the early leavers (dark blue squares) are typically found more concentrated in a region between 0.5 and 1.5 R$_{200}$ from the cluster centre. For the few remainer satellites that remain within their hosts after passing the cluster core (red diamonds), they tend to be found no further out than approximately 1.5~R$_{200}$. This is likely because their host halos are more massive and suffer stronger dynamical friction, and so are unable to backsplash out as far after passing the cluster core. Meanwhile, those satellites which leave can backsplash out to a much larger radius in the cluster. This is similar to the example given in Figure \ref{fig:fig4}, where group break up can scatter leavers out to a large clustercentric radius.

We note that the majority of the remainers that have yet to reach pericentre (orange pentagon symbols) are concentrated in one region simply because they are all members of a single, very populous host that is just entering the cluster for the first time. We also see some cases where the host ends up closer to the cluster centre than its leaver satellites, thus we checked if dynamical friction is playing a role in dragging the host into the cluster centre, if the host is much more massive than the satellites. However, we find it difficult to find clear evidence for this, perhaps because the ratio of host to cluster mass is always quite small ($m_{host}/m_{cl}$ < 0.1) in our somewhat relaxed clusters so dynamical friction is rather ineffective. We also test if dynamical friction acting on the host halo might cause those satellites that leave to move on different orbits after escaping (green points are beyond red ones; right panel). However, once again, we did not see any clear difference in their location in the phase-space at $z=0$ as a function of their host-to-cluster mass ratio.

We emphasise that the leaver satellites are widely distributed across phase-space at $z=0$. This further highlights that the mixing of leavers with the rest of the cluster population occurs quickly and effectively. As discussed in Section \ref{sec:when do sate...}, this likely makes identification of the galaxies that were previously in groups very challenging.

\subsection{Satellite accretion before and after entering the cluster }
\label{sec:Satellite accretion...}

\begin{figure}
	\includegraphics[width=\columnwidth]{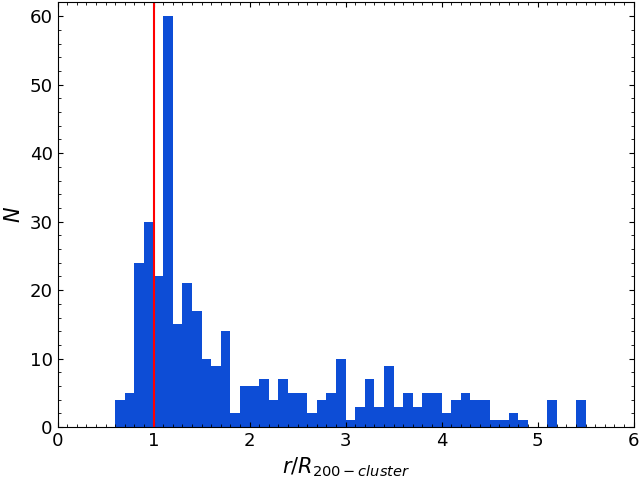}
    \caption{Distribution of the host clustercentric distances whenever they accrete a new satellite. The vertical red line indicates a clustercentric distance of 1R$_{200}$. }
    \label{fig:fig13}
\end{figure}

In this section, we check to see at what distance from the cluster hosts accrete new satellites (i.e. when they become bound for the first time according to our binding criterion). This is shown in Figure \ref{fig:fig13}. The first thing to note is that there is a prominent peak close to the R$_{200}$ radius of the cluster, specifically at 1.1-1.2~R$_{200}$. This indicates that there is an increase in the rate of satellite accretions happening in the outskirts of the cluster. At first glance, this appears to indicate that hosts are sweeping up galaxies in that region, in the manner described by \citet{Vijayaraghavan2013}. We checked if the swept up galaxies were previously in the cluster (i.e. backsplashed galaxies). However, we find this is not the case, and that the swept up galaxies are also falling into the cluster, alongside the host.

It is not fully clear why the accretion peaks so suddenly near the cluster R$_{200}$ radius. One possibility is that there may be more low mass halos available for accretion near the cluster, as the over dense region from which the cluster forms is more likely to form such halos in recent times. We find, however, that many of the halos formed early on, and seem to be falling in alongside their future hosts. Alternatively, it may simply be that the higher density of galaxies converging on the cluster, combined with compressive tidal forces can temporarily enhance the merger rate at this radius from the cluster. We defer a more detailed analysis to a follow-up study.

Another interesting result that can be seen in Figure \ref{fig:fig13} is when the hosts stop accreting their satellites (i.e. the bars on the left side of the vertical line). As can be seen, accretion is quickly reduced once the host is inside the cluster, and comes to a complete halt inside of a critical distance. This happens at $r/$R$_{200}$ $\sim$ 0.8 and at $r/$R$_{200}$ $\sim$ 0.65 for the two clusters, respectively. As noted previously, once a galaxy group is inside the cluster's potential, the most loosely bound satellites can be removed. At this point, it is likely that it becomes very challenging to accrete nearby satellites that were previously infalling with the host, and might otherwise have been accreted in the absence of the cluster tides. And, of course, the high velocity dispersion of galaxies in the cluster makes a merger with a previously unassociated cluster member extremely unlikely.

\subsection{Investigating the destroyed satellites}
\label{sec:delving more...}

\begin{figure}
	\includegraphics[width=\columnwidth]{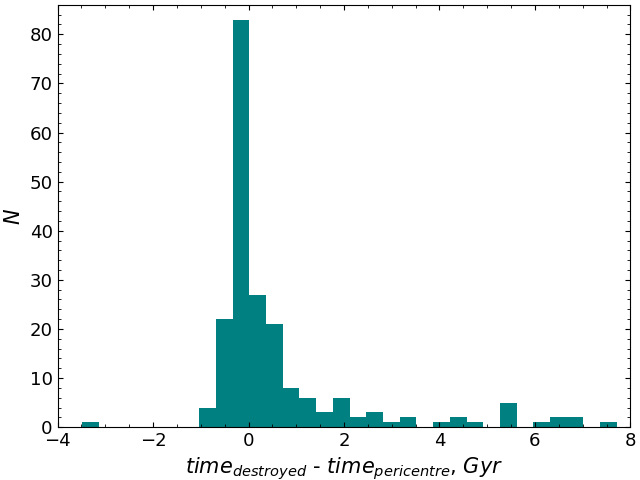}
    \caption{Distribution of the time, with respect to the moment of (first) pericentric passage of the host within the cluster, at which a satellite becomes destroyed.}
    \label{fig:fig14}
\end{figure}

To complement our analysis, in this section we take a closer look at the destroyed satellites \footnote{Note that some satellite hosts are also destroyed. However, this is rare ($\sim$ 13 $\%$ of the total host sample) and none contain satellites when they are destroyed.} as defined in Section \ref{sec:outcome of sat...}. In Figure \ref{fig:fig14}, we show the time at which they become destroyed relative to the time of first pericentric passage of their host in the cluster. As we saw previously, this is a key moment in the halos' evolution (see section \ref{sec:when do sate...}). From the figure we note that most of the satellites are destroyed during or after the pericentric passage, within a broad time window centred on that moment. For example, from the distribution we calculate that $\sim$42\% are destroyed in the 0.5 Gyr period after the first passage. Only around $\sim$5\% are destroyed 5 Gyr or more afterwards. In fact, many of these may survive several passages more in the cluster before being destroyed. On the other hand, the rest of the destroyed satellites (around 30\%) are destroyed just 1~Gyr before the first pericentre passage. Summing up, we find that almost 70\% of all those destroyed are destroyed by 0.5 Gyr after the first pericentric passage of the host. Note that we must exclude a small number of destroyed satellites (9) in hosts that are on first infall, but have not had a pericentric passage by $z=0$.

The fact that satellites are typically destroyed in the proximity of the cluster core supports our earlier statements on the importance of the cluster tidal forces at pericentre for dismantling substructure, as we saw previously with the leaver satellites. In this case we see that the cluster tides are capable of destroying satellites altogether. One caveat here is related to possible numerical effects and limitations of the simulation. We cannot rule out that some destroyed halos may be caused by broken merger trees. Fortunately, our merger tree building is quite robust in that it can survive the loss of a halo for one snapshot. If a halo is lost for more than one snapshot, however, then the tree will break. In any case we emphasise that these simulations have very high spatial and mass resolution which will help significantly to reduce numerical effects.

\section{Discussion}
\label{sec:discussion}

Galaxies that fall into the cluster are not doing so in isolation. Many fall in as part of a larger host halo. Previous studies \citep [e.g.][]{McGee2009,DeLucia12} find that as many as one third to one half of galaxies in clusters today were within groups previously. However, note that rather than considering groups (where the halo mass is typically considered to be higher than 10$^{13}$~M$_\odot$), we have considered ``hosts'', where there is no constraint on their halo mass. Also, our approach for defining the satellites of hosts is based on the specific energy criteria of \citet{Han2018}, which differs from the definition in \cite{McGee2009} and \cite{DeLucia12}. Nevertheless, they will contribute in much the same way to the pre-processed galaxy population in clusters. 

We find that 92.2\% of the satellites that have passed pericentre will be defined as leavers by $z=0$, and most of them leave quite shortly after pericentric passage, usually within 0.5 Gyr. This clearly illustrates the high efficiency with which the first pericentre passage disassembles substructure. We find that the satellites leave their hosts with quite high velocities and so, in many cases, they are quite far from their original hosts by $z=0$, with distances more than 10 times their host's current R$_{200}$ radius. This means they will be difficult to associate with their original host, and likely difficult to disseminate from the rest of the cluster population. 

Although, this may make recognising cluster members that were previously in a host more difficult, it also provides a potentially useful tool for determining where substructure is likely to be along its orbit through the cluster. Very few satellites ($\sim$7.8\%) remain bound to their host at $z=0$ if they passed first pericentre. Those that do remain bound tend to have a compact distribution within their host at $z=0$. Thus, if we observe substructure within a cluster such as a group of galaxies, it is probable that this substructure has only recently entered the cluster and has yet to reach first pericentre, which typically takes less than $\sim$1.5~Gyr. The probability that the substructure is on first infall rises even further if it is observed to be rather extended, with galaxy members reaching out beyond the group R$_{200}$ radius. We also do not see remainer satellites beyond roughly one R$_{200}$ radius of the cluster, perhaps due to the actions of dynamical friction on their groups. Therefore substructure visible beyond the cluster R$_{200}$ radius may be more likely on first infall into the cluster.

 We do not find strong evidence that satellites leave simply because their host's mass is reduced. Instead, we find that the main cause behind removal of a host's satellites is the tides of the cluster potential. Frequently satellites leave their host even before reaching the pericentre of their orbit, supporting our assertion that the cluster tides are responsible. Tidal shocking, which results from the rapid change in the tidal fields when a galaxy passes pericentre, could cause some satellites to leave, but this is only possible shortly after a host has passed pericentre. It seems, therefore, that the cluster tides provide a better explanation for the majority of our leaver sample.

It has previously been shown that pre-processing can contribute significantly to the tidal mass loss of galaxy dark matter halos \citep[e.g.][]{Joshi2017,Han2018}. Therefore we decided to study the tidal mass loss of satellites that are in the cluster but also suffer additional mass loss from their groups. We find a weak signal that this is occurring but the reason it is weak is likely also related to the efficiency by which substructure is disassembled by the cluster tides at first pericentric passage. Most satellites in hosts do not lose much additional mass within the cluster because they simply do not spend much time in their hosts before reaching the first pericentre and being removed from the host.

Related to this, we also studied where the destruction of our satellite halos occurs. This occurs very commonly near cluster pericentre. This further highlights how the cluster tides ---not the host's tides---dominate the tidal mass loss of our sample, at least during the period they spend within the cluster.

We also find supporting evidence for the suggestion in \cite{Vijayaraghavan2013} that groups which are merging with the cluster can sweep up some galaxies in the cluster outskirts. There is a prominent peak in the satellite accretion rate at around 1-1.2~R$_{200}$ from the cluster centre. In fact, around 21\% of the total satellite accretions occurs here. One consequence of this is that some of those satellites which join the cluster in a host may have only spent a very short time in the host, which potentially could reduce the significance of pre-processing for them.

As mentioned in section \ref{sec: causes: separation...}, we found that the separation distance between a group member and its group at first pericentre was a key factor in deciding if group members would be stripped from their group by the cluster. However, we believe an improvement on our study would be to use a more sophisticated treatment of the tidal radius, such as that described in \cite{Read2006}, where it was applied to star clusters. Applying this treatment to our case would allow the tidal radius to depend not only on the potential of the cluster, and the potential of the host, but also the orbit of the host and the orbit of the satellite within the host.

Another limitation of our study, in particular for the fraction of tidal mass loss of the satellites and their hosts, is that our simulations are dark matter only. Therefore we cannot directly measure the effect that tidal stripping might have on their baryonic component, which of course is the only component directly visible to us. However, as shown by \cite{Smith2016}, the relationship between dark matter tidal stripping and stellar tidal stripping in hydrodynamical cosmological simulations is actually quite well constrained: only halos that have lost more than $\sim$80\% of their dark matter halo are likely to suffer any stellar stripping. Then, as the remaining $\sim20\%$ of the dark matter is stripped, the stellar fraction reduces rapidly towards 0$\%$. Hence, only those halos that we consider to be very heavily tidally stripped would have suffered any significant stellar stripping. For instance, for the lowest mass halos of our sample, we can assume that the stellar mass loss would be negligible until the halo disappears below the resolution limit of our simulation. But for halos with a mass $2\times10^9$ M$_\odot$ (ten times more massive than our lowest mass halos), thanks to the high mass resolution of the simulations, we can follow the dark matter mass loss down to 3\% of their original value, which implies that most of the stars will have been stripped (for example, see Figure 3 of \citealp{Smith2016}). 
Therefore, our knowledge of the amount of stellar stripping is dependent on how far above the detection limit a halo's mass is. But with regards to stripping of gas by hydrodynamical mechanisms, and consequently the changes in galaxy star formation rates, we can provide little constraints due to the lack of hydrodynamics in these simulations.

\section{Conclusions}
\label{sec:conclusions}

Several theoretical studies suggest significant fractions of galaxies in clusters have spent time in galaxy group environments prior to infall into the cluster, and have attempted to study the impact this might have for galaxy evolution \citep[e.g.][]{McGee2009,DeLucia12,Han2018}, a phenomenon known as ``pre-processing''. It therefore follows that some cluster members may face the combined environmental influence of the cluster and their host, a process known as ``post-processing''. Motivated by this concept, we studied a set of high resolution N-body cosmological zoom simulations of galaxy clusters to attempt to better understand the outcome of galaxies which enter clusters as satellites of another host system, such as a group. The key results of this study may be summarised as follows.
\begin{itemize}

    \item We find that hosts which pass pericentre have their satellites efficiently broken up by the cluster tides. 92.2\% of such satellites leave their host, and most leave quite shortly (within 0.5~Gyr) after pericentric passage.

    \item The primary mechanism breaking up the substructure is the cluster tides, and not tidal shocking, as is evident from the fact that the majority of leavers exit their hosts even before reaching pericentre. Hosts do not have to lose a significant amount of mass for their satellites to be removed.
    
    \item The efficient break-up of substructure on first passage, and the fact that those few satellites which do remain bound tend to have a compact distribution within their hosts at $z=0$, may provide a useful tool for assessing where in its orbit through the cluster the substructure is today. Most substructure observed in clusters was likely only recently accreted and has yet to pass first pericentre. The probability that a host is on first infall is even greater if the satellites extend beyond the R$_{200}$ radius of their host.
    
    \item Satellites that leave their hosts do so at quite high velocities (typically $\sim$150~km/s) and so rapidly separate from their former hosts, which will likely make it more difficult to identify them from amongst the general cluster population.
    
    \item We find there is only a weak increase in the tidal mass loss of satellites due to their host's tides during their time in the cluster. But this is likely because they are quickly separated from their hosts. There is also a strong peak in satellite destruction near the time of pericentric passage, which further confirms that the cluster tides dominate their mass loss after they fall into the cluster.

    \item Outside of the cluster, we see a prominent peak in the ability of hosts to accrete satellites near the cluster R$_{200}$ radius. But, once inside the cluster, the ability for hosts to accrete new satellites is rapidly decreased, and completely halts inside of $\sim$0.6$-$0.8 R${_{\rm{200}}}$.
\end{itemize}

Many observed clusters show clear evidence for containing rich substructure from groups to subclusters. Thus, there is direct evidence that galaxies infalling into a cluster do not always do so in isolation, and could be previously influenced by the group environment. Our results provide some insight into what occurs to those satellite galaxies that enter clusters in a host system, the fate of substructure within the cluster, and may help us to understand when the substructure entered the cluster and thus give better constraints on the cluster's merger history.

In the future, we will extend the analysis to simulation data for more clusters, with a greater variation in cluster mass, and to give better number statistics. We will also consider the impacts of post-processing on baryons of these satellites, by using hydrodynamical cosmological simulations of clusters. Furthermore, since we only considered relaxed clusters so far, it would also be worthwhile to study how our results depend on the dynamical state of the cluster. Given that our clusters are dynamically relaxed, we do not have particularly massive groups in our sample, which may have reduced the significance of dynamical friction on our results. Extending our analysis to also include these cases may potentially aid us in developing approaches to locate cluster members that were previously in a group.

\section*{Acknowledgements}

We thank the anonymous referee for several comments that have helped to improve the paper. R.P. acknowledges the financial support from the European Union's Horizon 2020 research and innovation programme under the Marie Sk\l{}odowska-Curie grant agreement No. 721463 to the SUNDIAL ITN network. N.C.C. acknowledges the financial support from CONICYT PFCHA/DOCTORADO BECAS CHILE/2016 - 72170347.


\bibliographystyle{mnras}

\bibliography{references} 



\bsp	
\label{lastpage}
\end{document}